\documentclass[aps,twocolumn,superscriptaddress,amsmath,amssymb]{revtex4-1}
\usepackage{mathtools}
\DeclarePairedDelimiter\abs{\lvert}{\rvert}
\DeclarePairedDelimiter\norm{\lVert}{\rVert}
\DeclareMathOperator{\Tr}{Tr}
\DeclareMathOperator*{\argmin}{argmin}
\DeclareMathOperator*{\mpsprod}{\overset{\bullet}{\bigotimes}}
\usepackage[colorlinks,linkcolor=blue,citecolor=blue]{hyperref}

\newcommand{\Hbb}{\mathbb{H}}
\newcommand{\Hcal}{\hat{\mathcal{H}}}
\newcommand{\Scal}{\mathcal{S}}
\newcommand{\Pcal}{\mathcal{P}}
\newcommand{\Tcal}{\mathcal{T}}
\newcommand{\Jcal}{\mathcal{J}}

\newcommand{\Rrm}{\mathrm{R}}
\newcommand{\Lrm}{\mathrm{L}}
\newcommand{\irm}{\mathrm{i}}
\newcommand{\Sbf}{\hat{\mathbf{S}}}
\newcommand{\sbf}{\hat{\mathbf{s}}}
\newcommand{\vbf}{\mathbf{v}}
\newcommand{\ubf}{\mathbf{u}}
\newcommand{\hbf}{\mathbf{h}}

\newcommand{\eff}{\mathrm{eff}}
\newcommand{\loc}{\mathrm{loc}}
\newcommand{\sums}{\sum_{\{s\}}}
\newcommand{\ulin}{\underline{\hphantom{A}}}

\newcommand{\bmqty}[1]{\begin{bmatrix}#1\end{bmatrix}}
\newcommand{\sbmqty}[1]{\scriptsize{\bmqty{#1}}}
\newcommand{\ket}[1]{\left\lvert\smash{#1}\right\rangle}

\newcommand{\bra}[1]{\left\langle\smash{#1}\right\rvert}

\newcommand{\braket}[2]{\left\langle\smash{#1}\middle\vert\smash{#2}\right\rangle}
\newcommand{\Braket}[2]{\left\langle{#1}\middle\vert{#2}\right\rangle}
\newcommand{\mel}[3]{\vphantom{#1#2#3}\left\langle\smash{#1}\middle\vert\smash{#2}\middle\vert\smash{#3}\right\rangle}

\newcommand{\Mel}[3]{\left\langle{#1}\middle\vert{#2}\middle\vert{#3}\right\rangle}

\begin{document}

\title{Characterization of localized effective spins in gapped quantum spin chains}

\author{Hayate Nakano}
\email{hnakano@exa.phys.s.u-tokyo.ac.jp}
\affiliation{Department of Physics, University of Tokyo, 7-3-1 Hongo, Tokyo 113-0033, Japan}

\author{Seiji Miyashita}
\email{miyashita@phys.s.u-tokyo.ac.jp}
\affiliation{The Physical Society of Japan, 2-31-22 Yushima, Tokyo 113-0033, Japan}
\affiliation{Institute for Solid State Physics, University of Tokyo, 5-1-5 Kashiwanoha, Kashiwa 277-8581, Japan}
\affiliation{Elements Strategy Initiative Center for Magnetic Materials, National Institute for Materials Science, 1-2-1 Sengen, Tsukuba 305-0047, Japan}

\date{\today}

\begin{abstract}
We study properties of localized effective spins induced in gapped quantum spin chains by local inhomogeneities of the lattice. 
As a prototype, we study effective spins induced in impunity sites doped AKLT model by constructing the exact ground state in a matrix product state (MPS) form. 
We characterize their responses to external fields by studying  an extended Zeeman interaction. 
We also study the antiferromagnetic bond-alternating Heisenberg chain with defect structures.
For this model, an MPS representation similar to that for the AKLT model, ``a uniform MPS with windows,'' is constructed, and it gives a good approximation of the ground state.
We discuss the trade-off relation between the window length and the precision of the MPS ansatz.
The effective exchange interaction between the induced spins is also investigated by using this representation.
\end{abstract}

\maketitle

\section{Introduction}\label{sec:Intro}

Collective motions in quantum many-body systems are one of the most exciting topics in quantum dynamics, which give the basis of recently developing quantum information techniques~\cite{Brennen2008-xz,Miyake2010-xo,Bartlett2010-xe,Meier2003-rm,Srinivasa2007-in,Liu2013-fc}. 
As a typical example of such collective phenomena in quantum systems, it has been well studied that localized effective spins are induced in gapped quantum spin systems.
Such structures appear at local inhomogeneities in lattices, e.g., edges, impurity spins, inhomogeneities of interactions, etc.

For example, edges and impurities in the $S=1$ antiferromagnetic Heisenberg chain have been studied extensively~\cite{Kaburagi1994-qq,Sorensen1995-td,Ramirez1994-xv,Wang1996-xl,Batista1999-se}.
Moreover, localized spin moments at the inhomogeneous structure are pointed out in several systems~\cite{Nishino2000-xc,Nishino2000-np}. 

Recently, the coherent dynamics of such localized magnetic structure has been measured in experiments. 
For example, Bertaina, \textit{et al.}~\cite{Bertaina2014-ww} measured Rabi oscillations of the localized spins in $(\text{TMTTF})_2\mathrm{PF_6}$, which was modeled by the antiferromagnetic bond-alternating Heisenberg chain (ABAHC). 
They also discussed the effect of the localized spins on the ESR spectrum and proposed possible use of the magnetic structure as a spin qubit. 

Under these circumstances, the theoretical analysis of such localized effective spins becomes more important.
In the present paper, we characterize such effective spins as a collective mode in gapped systems by making use of the matrix product state (MPS) representation~\cite{White1992-oy,Schollwock2005-kh,Schollwock2011-cw,Perez-Garcia2007-ed} and study their coherent responses to the external field. 

As a prototype, we first study the Affleck-Kennedy-Lieb-Tasaki (AKLT) model~\cite{Affleck1987-aj,Affleck1988-si}. 
The AKLT model is a frustration-free spin model, and the exact uniform MPS representation of the ground state exists~\cite{Fannes1992-id,Klumper1993-zk}. 
We dope $S =3/2$ impurity spins into the AKLT model and introduce interactions around them with projection operators in order not to break the frustration-free property.
Then, the ground state exhibits $S=1/2$ effective spin structures.
By replacing the tensors at the impurity sites, we can construct the exact MPS representation of the effective spin states.
We call such MPS structure ``a uniform MPS with windows.''
By making use of this MPS representation, we study responses to an external magnetic field and propose a way of independent manipulation of two distinct systems (qubits) in two effective spins systems.
We also point out such manipulation is not possible for more than two spins.

As a more realistic model, we study the ABAHC, which has the gapped ground state and inhomogeneities cause localized effective spins. 
Because this is not a frustration-free model, the discussions of the AKLT model are not fully applicable. 
However, MPS based analyses are still useful in this case.
We studied the ABAHC with weak-weak bond defects and investigated the interaction between two effective spins induced by the defects.
We obtained the asymptotic behavior of the interaction strength as a function of the separation by using the MPS representation with the windows.
We also discuss the trade-off relation between the precision and the window length, both numerically and analytically.

The present paper is organized as follows. 
In Sec. \ref{sec:AKLT}, we study MPS for the AKLT model with impurities as a prototype, and in Sec. \ref{sec:ABAHC}, MPS for the ABAHC are given. 
Summary and discussion are given in Sec. \ref{sec:Summary}.

\section{Localized spin structure in the AKLT model}\label{sec:AKLT}

The AKLT model is an $S=1$ antiferromagnetic quantum spin chain described by the Hamiltonian
\begin{align}
    \Hcal_{\mathrm{AKLT}} = \sum_{i} \left[{\Sbf}_i \cdot {\Sbf}_{i+1} + \frac{1}{3}\left({\Sbf}_i \cdot {\Sbf}_{i+1}\right)^2 + \frac{2}{3}\right].
\end{align}
${\Sbf}_i \cdot {\Sbf}_{i+1} + \frac{1}{3}({\Sbf}_i \cdot {\Sbf}_{i+1})^2 + \frac{2}{3}$ is proportional to $\hat{\Pcal}_{i,i+1}^{S_{\mathrm{tot}}=2}$, which is defined as the projection operator onto the spin $2$ subspace of $\Hbb_i^{S=1} \otimes \Hbb_{i+1}^{S=1}$. 
Here, $\Hbb_i$ denotes the local Hilbert space at site $i$.

\begin{figure}
\centering
\includegraphics{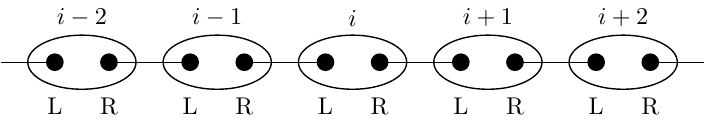}
\caption{The schematic picture of the VBS state. Black dots denote $s_{\Lrm,\Rrm}$ and white circles denote the symmetrization operator $\Scal$.}\label{fig:vbs}
\includegraphics{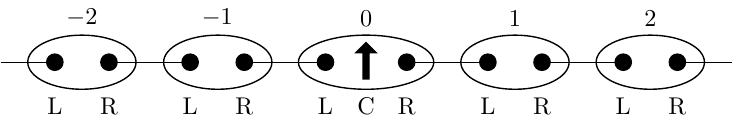}
\caption{The effective spin state with a doped $S=\frac{3}{2}$ spin.}\label{fig:vbs32}
\end{figure}

For a chain with periodic boundary condition, the ground state is given as follows: \begin{align}
    \left(\bigotimes_i \Scal_i \right)\ket{\mathrm{dimer}} = \Tr\left[\mpsprod_i \bmqty{-\frac{1}{2}\ket{0_i} && \frac{1}{\sqrt{2}}\ket{+_i} \\ -\frac{1}{\sqrt{2}}\ket{-_i} && \frac{1}{2}\ket{0_i}}\right],\label{eq:oldakltmps}
\end{align}
where the operator $\dot{\otimes}$ acts as $(X\dot{\otimes}Y)_{i,j} = \sum_k X_{i,k}\otimes Y_{k,j}$ and $\ket{\mathrm{dimer}}$ denotes a dimer state of virtual $S=1/2$ spins
\begin{align}
    \ket{\mathrm{dimer}} &= \bigotimes_i \frac{\ket{\uparrow_{i,\Rrm}}\ket{\downarrow_{i+1,\Lrm}}-\ket{\downarrow_{i,\Rrm}}\ket{\uparrow_{i+1,\Lrm}}}{\sqrt{2}} \nonumber\\&= \Tr \left[\mpsprod_i \frac{1}{\sqrt{2}}\bmqty{-\ket{\uparrow_{i,\Lrm}}\ket{\downarrow_{i,\Rrm}} & \ket{\uparrow_{i,\Lrm}}\ket{\uparrow_{i,\Rrm}} \\ -\ket{\downarrow_{i,\Lrm}}\ket{\downarrow_{i,\Rrm}} & \ket{\downarrow_{i,\Lrm}}\ket{\uparrow_{i,\Rrm}}}\right].\label{eq:dimerdef}
\end{align}
The operator $\Scal^{S=1}$ is a symmetrization operator, which makes two virtual $S=1/2$ spins into a $S=1$ spin as
\begin{align}
\Scal^{S=1}\ket{\tfrac{1}{2};s_\Lrm}\ket{\tfrac{1}{2};s_\Rrm} = \frac{1}{\sqrt{\binom{2}{1+s_\Lrm+s_\Rrm}}}\ket{1;s_\Lrm+s_\Rrm},\label{eq:symmet}
\end{align}
where $\binom{n}{m}$ denotes the combination number.
This ground state is called a valence bond solid (VBS) state and often illustrated in a schematical picture depicted in Fig.~\ref{fig:vbs}.

This state is written in the conventional form of MPS:
\begin{align}
    \sums \Tr \bigl(\prod_i A^{s_i}\bigl) \ket{\{s\}}
\end{align}
with the tensor with theree indices
\begin{align}
    A^{+} &= \sqrt{\frac{2}{3}} \sigma^+ ,\quad A^0 = -\sqrt{\frac{1}{3}} \sigma^z,\quad A^- = -\sqrt{\frac{2}{3}}\sigma^- .\label{eq:defA}
\end{align}
Here, $\ket{\{s\}}$ denotes the basis of the system $\ket{s_1,s_2,\dots,s_N}(s_i = +,0,-)$ and $\sigma$ denotes the Pauli matrices. 
The coefficients of \eqref{eq:defA} are introduced into $A$ to make the state normalized in the thermodynamic limit.

For a chain with open boundary condition, the ground state is obtained by applying the symmetrization operators \eqref{eq:symmet} on the following state
\begin{align}
    \ket{s_{1,\Lrm}}\otimes \left(\bigotimes_{i=1}^{N-1} \frac{\ket{\uparrow_{i,\Rrm}}\ket{\downarrow_{i+1,\Lrm}}-\ket{\downarrow_{i,\Rrm}}\ket{\uparrow_{i+1,\Lrm}}}{\sqrt{2}}\right)\otimes \ket{s_{N,\Rrm}},
\end{align}
instead of the periodic dimer state \eqref{eq:dimerdef}.
Because of the edge spins $s_{1,\Lrm},s_{N,\Rrm}$, the ground state is four-fold degenerate.
We mention that these $S=1/2$ spin degrees of freedom are localized but not strictly localized around the edges.
To make it clear, let us consider the magnetization profile in the case of $s_{1,\Lrm} = s_{N,\Rrm} = \ \uparrow$ and $N \rightarrow \infty$.
Around the left edge, the profile is given by
\begin{align}
    \left<S_i^z\right> = -2\left( -\frac{1}{3}\right)^i,\quad i = 1,2,\ldots.
\end{align}
Since $\sum_{i=1}^\infty -2(-1/3)^{i} = 1/2$, this structure can be regarded as a localized $S=1/2$ spin originating from $s_{1,\Lrm}$.

\subsection{AKLT model with impurity spins}

Here, we consider the AKLT model with a doped $S = 3/2$ spin, which induces an $S=1/2$ localized spin structure. 
We tune the interactions around the doped spins to make the ground state exactly representable in the MPS form.
The constructed Hamiltonian actiong on $(\bigotimes_{i<0}\mathbb{H}^{S=1}_i)\otimes \mathbb{H}^{S=\frac{3}{2}}_0\otimes (\bigotimes_{i>0}\mathbb{H}^{S=1}_i)$ is
\begin{align}
    &\Hcal = \sum_{\substack{i<-1\\\mathrm{or}\,1\leq i}}\left[\Sbf_i \cdot \Sbf_{i+1} + \frac{1}{3}(\Sbf_i \cdot \Sbf_{i+1})^2 + \frac{2}{3}\right]+ \left[\Sbf_{-1} \cdot \sbf_{0} \vphantom{\frac{2}{3}} + \right. \nonumber\\ &\left. \frac{2}{7}(\Sbf_{-1} \cdot \sbf_{0})^2 + \frac{5}{7}\right] + \left[\sbf_{0} \cdot \Sbf_{1} + \frac{2}{7}(\sbf_0 \cdot \Sbf_{1})^2 + \frac{5}{7}\right],
\end{align}
where $\sbf$ denotes an $S = 3/2$ spin operator. Since the interaction around the impurity is also proportional to the projection operator, i.e.,
\begin{align}
    \hat{\Pcal}_{i,i+1}^{S_{\mathrm{tot}}= \frac{5}{2} } \propto
 \sbf_i \cdot \Sbf_{i+1} + \frac{2}{7} (\sbf_i \cdot \Sbf_{i+1})^2 +\frac{5}{7},
\end{align}
the ground state of this Hamiltonian can be constructed in the same way as that of the uniform AKLT model. 
The ground state is schematically expressed in Fig.~\ref{fig:vbs32}. 
Let us briefly illustrate how to construct the MPS representation of this state. 
First, we construct a product state of the dimer state \eqref{eq:dimerdef} and an extra $S=1/2$ spin state $\ket{s_
{0,\mathrm{C}}}$,

\begin{align}
    &\ket{\mathrm{dimer}}\otimes \ket{s_{0,\mathrm{C}}} \propto \nonumber\\ 
    &\begin{alignedat}{2}
    \Tr \left[\vphantom{\mpsprod_{i}}\right.&\left(\mpsprod_{i<0}   \bmqty{-\ket{\uparrow_{i,\Lrm}}\ket{\downarrow_{i,\Rrm}}  && \ket{\uparrow_{i,\Lrm}}\ket{\uparrow_{i,\Rrm}} \\ -\ket{\downarrow_{i,\Lrm}}\ket{\downarrow_{i,\Rrm}}  && \ket{\downarrow_{i,\Lrm}}\ket{\uparrow_{i,\Rrm}}}\right)\dot{\otimes}& &\\& \bmqty{-\ket{\uparrow_{0,\Lrm}}\ket{s_{0,\mathrm{C}}}\ket{\downarrow_{0,\Rrm}} && \ket{\uparrow_{0,\Lrm}}\ket{s_{0,\mathrm{C}}}\ket{\uparrow_{0,\Rrm}} \\ -\ket{\downarrow_{0,\Lrm}}\ket{s_{0,\mathrm{C}}}\ket{\downarrow_{0,\Rrm}} && \ket{\downarrow_{0,\Lrm}}\ket{s_{0,\mathrm{C}}}\ket{\uparrow_{0,\Rrm}}} \dot{\otimes} & &\\& \left(\mpsprod_{0<i} \bmqty{-\ket{\uparrow_{i,\Lrm}}\ket{\downarrow_{i,\Rrm}}  && \ket{\uparrow_{i,\Lrm}}\ket{\uparrow_{i,\Rrm}} \\ -\ket{\downarrow_{i,\Lrm}}\ket{\downarrow_{i,\Rrm}}  && \ket{\downarrow_{i,\Lrm}}\ket{\uparrow_{i,\Rrm}}}\right)&\left.\vphantom{\mpsprod_i}\right]& .
    \end{alignedat}
\end{align}

The symmetrization operator acting on three $S=1/2$ spins is now given by
\begin{align}
    &\Scal^{S=\frac{3}{2}}\ket{\tfrac{1}{2};s_{\Lrm}}\ket{\tfrac{1}{2};s_{\mathrm{C}}}\ket{\tfrac{1}{2};s_{\Rrm}} = \nonumber\\ &\frac{1}{\sqrt{\binom{3}{\frac{3}{2}+s_{\Lrm}+s_{\mathrm{C}}+s_{\Rrm}}}}\ket{\tfrac{3}{2};s_{\Lrm}+s_{\mathrm{C}}+s_{\Rrm}}.    
\end{align}

By applying $(\bigotimes_{i<0} \Scal_i^{S=1})\otimes \Scal_0^{S=\frac{3}{2}}\otimes(\bigotimes_{0<i}\Scal_i^{S=1})$ to $\ket{\mathrm{dimer}}\otimes \ket{s_{0,\mathrm{C}}}$, we obtain the ground state
\begin{align}
    \ket{\sigma^{\loc}} = \sums\Tr \bigl[\bigl(\ \prod_{\mathclap{-N/2<i<0}}\ A^{s_i}\bigl)B_{\sigma^{\loc}}^{s_0}\bigl(\ \prod_{\mathclap{0<i\leq N/2}}\ A^{s_i}\bigl)\bigl] \ket{\{s\}}, \label{eq:akltsand}
\end{align}
where $\sigma^{\loc}=\uparrow,\downarrow$ denotes the index of the $S=1/2$ localized spin corresponding to the unpaired spin $s_\mathrm{C}$, and non-zero elements of $B$ are defined by
\begin{align}
    B^{+\frac{3}{2}}_{\uparrow} =\sigma^+ ,\quad B^{+\frac{1}{2}}_{\uparrow} = -\sqrt{\frac{1}{3}}\sigma^z ,\quad B^{-\frac{1}{2}}_{\uparrow} = -\sqrt{\frac{1}{3}}\sigma^- \nonumber\\
    B^{+\frac{1}{2}}_{\downarrow} = \sqrt{\frac{1}{3}}\sigma^+ ,\quad  B^{-\frac{1}{2}}_{\downarrow} = -\sqrt{\frac{1}{3}}\sigma^z ,\quad  B^{-\frac{3}{2}}_{\downarrow} = -\sigma^-.\label{eq:defB}
\end{align}
From now, we consider only in the thermodynamic limit, i.e., $N\rightarrow\infty$ limit of \eqref{eq:akltsand}~\footnote{It is straightforward to generalize the discussions in this section for finite-size systems. But then, despite the formula becomes very complicated, the conclusion remains essentially unchanged.}. 
In this limit, the states are normalized, and the magnetization profiles of them are given by
\begin{align}
&\begin{aligned}
    \mel{\uparrow^\loc}{S^{(z)}_i}{\uparrow^\loc} &= \begin{cases}
    \frac{5}{6} & (i = 0)\\
    \frac{2}{3} \left(-\frac{1}{3}\right)^{|i|} & (i \neq 0)
    \end{cases} =: f(i)
\end{aligned}\\
    &\mel{\downarrow^\loc}{S^{(z)}_i}{\downarrow^\loc} = -f(i).
\end{align}
We note that $\sum_i f(i) = 1/2$ is satisfied.
Thus we succeeded to create effective $S=1/2$ spins represented by compact tensors by doping $S=3/2$ spins.
Here, it should be noted that we may construct the effective spin states by simply introducing $S=1/2$ spins~\cite{Kaburagi1994-qq,Sorensen1995-td}.
However, in this case, there is no compact representation since the projection method used above is not available.

\subsection{Response to magnetic field}

Now, we discuss a response of the effective spin structures to external magnetic fields described by the Hamiltonian $\Hcal'(t) = \sum_i \hbf_i(t) \cdot \Sbf_i$. For simplisity, hereafter we omit the argument $t$.
First, we consider the case of a uniform magnetic field $\Hcal' = \hbf \cdot (\sum_i  \Sbf_i)$. 
Since $[\Hcal,\Hcal']=0$ and $\Hcal\ket{\sigma^{\loc}} = 0$, the dynamics is bounded in the ground state subspace. 
The matrix representation of $\Hcal'$

\begin{align}
 \bmqty{\bra{\uparrow^\loc} \\ \bra{\downarrow^\loc}}\Hcal'\bmqty{\ket{\uparrow^\loc} && \ket{\downarrow^\loc}} = \frac{1}{2}\sum_{\alpha = x,y,z} h_\alpha \sigma^\alpha,
\end{align}
is the same as that of the Hamiltonian $\hbf \cdot \Sbf$ acting on a single free $S=1/2$ spin. 
Therefore, the response is the same as that of free $S =1/2$ spin.

In the case of non-uniform external fields, i.e., $\{\hbf_i\}$ is position-dependent, $\Hcal$ and $\Hcal'$ no longer commute. 
Here, we assume that the gap above the ground state is large, and the transition to the excited states is negligible.
Under this assumption, we study dynamics only in the ground states.
Then, the matrix representation of $\Hcal'$ is written as
\begin{align}
 \bmqty{\bra{\uparrow^\loc} \\ \bra{\downarrow^\loc}}\Hcal'\bmqty{\ket{\uparrow^\loc} && \ket{\downarrow^\loc}} = \frac{1}{2}\sum_{\alpha = x,y,z} h_{\alpha}^{\eff} \sigma^\alpha,
\end{align}
where we define effective magnetic fields as
\begin{align}
 h_{\alpha}^{\eff} = \frac{\sum_i f(i)h_{i,\alpha}}{\sum_i f(i)} = 2\sum_i f(i)h_{i,\alpha}.\label{eq:heffdef}
\end{align}
Thus, the response can be regarded again as the same as the free spin.

This observation indicates that the effective spin acts in the same way as long as the effective field is the same. 
Because of the one-to-many correspondence between $\hbf^{\eff}$ and $\{\hbf_i\}$, we can construct many different $\{\hbf_i\}$s which generate the same dynamics. 
The degrees of freedom of effective fields suggests the possibility to manipulate multiple effective spins independently by tuning the distribution $\{\hbf_i\}$.
Thus, in the following, we study the systems with multiple doped spins.

\subsection{MPS of multiple induced spins}

Now, we consider the Hamiltonian with multiple $S=3/2$ doped spins. 
By using the above-introduced tensor $B$, we can construct the ground state as
\begin{align}
    &\ket{\sigma^{\loc}_1,\dots,\sigma^{\loc}_{k}} = \nonumber\\ &\begin{alignedat}{2}
        \sums\Tr\bigl[ & & \bigl(\ \prod_{\mathclap{i<j_1}}\ A^{s_i}\bigl)B_{\sigma^{\loc}_1}^{s_{j_1}}\bigl(\ \prod_{\mathclap{j_1<i<j_2}}\ A^{s_i}\bigl)B_{\sigma^{\loc}_2}^{s_{j_2}}\dots &\\& &\dots B_{\sigma^{\loc}_k}^{s_{j_k}}\bigl(\ \prod_{\mathclap{j_k<i}}\ A^{s_i}\bigl) &\bigl] \ket{\{s\}}.
    \end{alignedat}\label{eq:multiaklt}
\end{align}
The ground state is $2^k$-fold degenerate, where $k$ is the number of doped spins. 

Here, we consider the case of $k=2$, and we fix the positions of doped spins as $j_1 = 0$ and $j_2 = L \geq 2$. 
The matrix elements of spin operators are given by
\begin{align}
    &\mel{\uparrow_1^\loc,\uparrow_2^\loc}{S^{(z)}_i}{\uparrow_1^\loc,\uparrow_2^\loc} = g_1(i) + g_2(i)\\
    &\mel{\uparrow_1^\loc,\uparrow_2^\loc}{S^{(+)}_i}{\uparrow_1^\loc,\downarrow_2^\loc} = g_2(i)
\end{align}
and so on, where
\begin{align}
 g_1(i) = \left(1 + \frac{1}{4}\delta_{i,L}\right)f(i),\quad g_2(i) = g_1(L-i).   
\end{align}
Now, we define $\ubf_2$ as
\begin{align}
    \ubf_2=\bmqty{\ket{\uparrow_1^\loc,\uparrow_2^\loc} && \ket{\uparrow_1^\loc,\downarrow_2^\loc} && \ket{\downarrow_1^\loc,\uparrow_2^\loc} && \ket{\downarrow_1^\loc,\downarrow_2^\loc}}.\label{eq:aklt4basis}
\end{align}
Then, the matrix elements of $\Hcal'$ is written as
\begin{align}
    &\ubf^\dagger_2 \Hcal' \ubf_2 \nonumber\\
    =&\frac{1}{2}\sbmqty{h_{z,1}^\eff + h_{z,2}^\eff && h_{x,2}^\eff - \irm h_{y,2}^\eff && h_{x,1}^\eff - \irm h_{y,1}^\eff && 0\\ h_{x,2}^\eff + \irm h_{y,2}^\eff && h_{z,1}^\eff - h_{z,2}^\eff && 0 && h_{x,1}^\eff - \irm h_{y,1}^\eff \\h_{x,1}^\eff + \irm h_{y,1}^\eff && 0 && -h_{z,1}^\eff + h_{z,2}^\eff && h_{x,2}^\eff - \irm h_{y,2}^\eff \\ 0 && h_{x,1}^\eff + \irm h_{y,1}^\eff && h_{x,2}^\eff + \irm h_{y,2}^\eff && -h_{z,1}^\eff - h_{z,2}^\eff}\nonumber\\
    =&\frac{1}{2}\sum_{j=1,2}\sum_{\alpha_j = x,y,z}\  h_{\alpha_j,j}^\eff\sigma^{\alpha_j}_j,\label{eq:22hmat}
\end{align}
where we define $h_{\alpha,j}^\eff = 2\sum_i g_j(i)h_{i,\alpha}$. 
Here it should be noted that the bases~\eqref{eq:aklt4basis} are not orthonormal and the Gram matrix is
\begin{align}
    G := \ubf_2^\dagger \ubf_2 = \sbmqty{1-\Delta_L && && && \\ && 1+\Delta_L && -2\Delta_L && \\ && -2\Delta_L && 1+\Delta_L && \\ && && && 1 - \Delta_L},
\end{align}
where $\Delta_L = (-\frac{1}{3})^{L+1}$. 
$\Delta_L$ can be regarded as a barometer of the overlap between the magnetization profiles of two effective spins. 
Because $G$ is different from the unit matrix $\mathbb{I}_{4\times4}$, the dynamics generated by $\Hcal'$ is different from that of two free $S=1/2$ spins.

To amend this difference, we introduce new basis $\{ \ket{\tilde{\sigma}_1,\tilde{\sigma}_2} \}$ by linear combinations of $\{ \ket{\sigma_1^\loc,\sigma_2^\loc} \}$ as
\begin{align}
    \tilde{\ubf}_2 =\ubf_2 \sqrt{G}^{-1} = \ubf_2 \sbmqty{\beta_+ + \beta_- && && && \\ && \beta_+ && \beta_- && \\ && \beta_- && \beta_+ && \\ && && && \beta_+ + \beta_-},
\end{align}
where
\begin{align}
\beta_\pm = \frac{1}{2}\left(\sqrt{\frac{1}{1 - \Delta_L}} \pm \sqrt{\frac{1}{1+3\Delta_L}}\right).
\end{align}
Then, the matrix elements of $\Hcal'$ for these new bases are given by the same form of \eqref{eq:22hmat} after redefining $h^\eff$ as $\frac{1}{\sqrt{1 - \Delta_L}}(\beta_+ h_{\alpha,1}^\eff + \beta_- h_{\alpha,2}^\eff) \rightarrow h^\eff_{\alpha,1}$.
Since the number of degrees of freedom of $\{\hbf_i\}$ is larger than that of $\{\hbf_k^\eff\}$, we can control $\hbf_{1}^\eff$ and $\hbf_{2}^\eff$ independently by tuning $\{\hbf_i\}$.
Thus, these new basis $\{ \ket{\tilde{\sigma}_1,\tilde{\sigma}_2} \}$ can be regarded as ``qubit'' states, which can be controlled independently by the external field.

\subsection{More than two spins} 

Now we study the case when the number of effective spins becomes larger than two. 

First, we consider the case of three spins. 
We found that it is impossible to properly define effective fields $h_{\alpha,k}^\eff=\sum_i f_k(i)h_{i,\alpha}$ and an orthonormal basis set $\{\ket{\tilde{\sigma}_1,\tilde{\sigma}_2,\tilde{\sigma}_3}\}$ which satisfies
\begin{align}
    \tilde{\ubf}_3^\dagger \Hcal' \tilde{\ubf}_3 = \sum_{j=1,2,3}\sum_{\alpha_j=x,y,z}h_{\alpha_j,j}^\eff\sigma^{\alpha_j}_j.\label{eq:3efspin}
\end{align}

In order to show this, we solve the generalized eigenvalue problem $\lambda G\vbf=H\vbf$ for \begin{align}
    H=\ubf_3^\dagger \left( \sum_i h_{i,z}S_i^z \right) \ubf_3,\quad G=\ubf_3^\dagger  \ubf_3.
\end{align}
If there exists a set of parameters satisfying \eqref{eq:3efspin}, the eigenvectors are independent of the choice of the configuration $\{h_{i,z}\}$.
To check whether such parameter sets exist or not, we generated random configurations and solved the eigenvalue problem numerically. 
Then, we found that different configurations make the eigenvectors different. 
Thus, we conclude that the ``qubit'' states which can be controlled independently are not possible for the case with three spins. 

This difference can be understood as a consequence of the scattering phenomena of the transfer matrices made of MPS (see Appendix \ref{sec:detailAKLT}).
In the case of more than two spins, as shown in \eqref{eq:mltsct}, multiple scattering more than two times causes peculiar matrix element in $H$.
Such scattering processes, which do not take place in the case of two spins, make the qualitative difference.

\section{Antiferromagnetic bond-alternating Heisenberg chain}\label{sec:ABAHC}

As mentioned in Introduction, effective spin structures are induced at local inhomogeneities in various kinds of gapped spin chains.
A typical example of such gapped chains is the spin-Peierls chain, modeled by the ABAHC.
Its Hamiltonian is given by
\begin{align}
    \Hcal_{\mathrm{ABAHC}} = \sum_i (1 + (-1)^i \delta)\Sbf_i \cdot \Sbf_{i+1},
\end{align}
where $\Sbf$ denotes an $S = 1/2$ spin operator.
We call the bond of the strength $1+\abs{\delta}$ ``strong bond'' and that of $1-\abs{\delta}$ ``weak bond''.
The ground state is thought to be in the same phase as the so-called dimer state. 
We can define a non-local string order parameter detecting the dimer order~\cite{Hida1992-um,Wang2013-jm}. 
These phases are regarded as the symmetry-protected topological phases named even-Haldane phase or odd-Haldane phase~\cite{Haghshenas2014-ml}.
Hereafter, we adopt the dimerization parameter as $\delta= 0.03$.
This value corresponds to the ESR experiment~\cite{Bertaina2014-ww} and is suitable to visualize the magnetization profile of the effective spin smoothly.

We use a uniform MPS
\begin{align}
    &\ket{\Psi(A)} = \sums \vbf_\Lrm^{\dagger}\bigl(\prod_{i\in \mathbb{Z}}A^{s_{2i},s_{2i+1}}\bigl) \vbf_\Rrm\ket{\{s\}}
\end{align}
to approximate the ground state. 
The tensor $A \in \mathbb{C}^{2^2\times D\times D}$ is defined for every two sites, and $\vbf_{\Lrm,\Rrm}$ are boundary vectors with $D$ complex elements.
$D$ is the bond dimension of $A$.
$A$ and $\vbf$ are chosen to satisfy the normalization $\Braket{\Psi(\bar{A})}{\Psi(A)} = 1$, where the overline denotes the complex conjugate.
To obtain the ground state, we optimize the tensor $A$ to minimize the energy $\Mel{\Psi(\bar{A})}{\Hcal_{\mathrm{ABAHC}}}{\Psi(A)}$.
In the present study, we use the VUMPS algorithm~\cite{Zauner-Stauber2018-bm} for this purpose. We prepare several normalized random tensors as initial states of the optimization and check that the optimized tensors are independent of the initial choices. It suggests that the obtained states are not trapped in local minimums of the energy.

The correlation length of the ground state is calculated by the transfer matrix $(\Tcal_A^A)_{(a,a'),(b,b')}:=\sum_s A_{a,b}^s \bar{A}^s_{a',b'}$. 
The eigenvalues of $\Tcal_A^A$, $\lambda_1,\lambda_2,\dots$, are sorted in descending order of their magnitude.
Because of the normalization, $\lambda_1$ is equal to 1.
We assume $\abs{\lambda_2}<1$, and the correlation length is defined by $\xi_\mathrm{bulk}=-1/\ln\abs{\lambda_2}$. 

For the present model, we found $D=130$ is large enough to study the qualitative characteristics of the effective spin structures, although the extrapolation to $D\rightarrow\infty$ gives a quantitative difference.
The correlation length of the $D=130$ MPS is $\xi_{\mathrm{bulk}} \approx 5.055$.
We also estimate the energy gap $\Delta E \approx 0.145$.

\subsection{Defects and effective spin structures}

Now, we study the ground state with a single defect.
It is known that the ground state of the ABAHC has an effective spin structure around the inhomogeneity~\cite{Nishino2000-xc,Nishino2000-np}. 
Here, we introduce a weak-weak bond defect into the ABAHC.
The lattice structure is schematically drawn as
\begin{align*}
\includegraphics{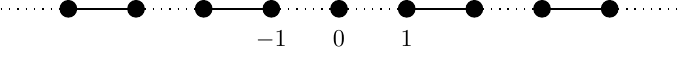}
\end{align*}
where solid and dotted lines denote the strong and weak bonds, respectively.

We calculate the ground state of this model in the MPS form.
We use the central $2N+1$ tensors, which we call ``window'' in the uniform MPS, to express the effect of the defect. 
Namely, the variational wave function is given by
\begin{align}
    &\ket{\Psi(A;\{B_{[i]}\}_{i=-N}^N)} =\nonumber\\ &\sums \vbf_\Lrm^{\dagger} \bigl(\prod_{n < -N} A^{s_{2n},s_{2n+1}}\bigl) B_{[-N]}^{s_{-2N},s_{-2N+1}} \dots B_{[-1]}^{s_{-2},s_{-1}}\times \nonumber\\ & B_{[0]}^{s_0}\dots B_{[N]}^{s_{2N-1},s_{2N}} \bigl(\prod_{N < n} A^{s_{2n-1},s_{2n}}\bigl) \vbf_\Rrm \ket{\{s\}},\label{eq:sMPS}
\end{align}
where $A$ is the tensor already calculated for the uniform model.
This form can be regarded as the generalization of \eqref{eq:akltsand}.
The TDVP algorithm~\cite{Haegeman2011-jb,Haegeman2013-oj,Haegeman2016-tu,Milsted2013-tq} was used for the optimization of $\{B_{[i]}\}$ (see also Appendix~\ref{sec:WL}). 
In the optimization, we apply a small magnetic field in the $z$-direction in order to break the degeneracy of the ground state.

\begin{figure}
\includegraphics[width=\linewidth]{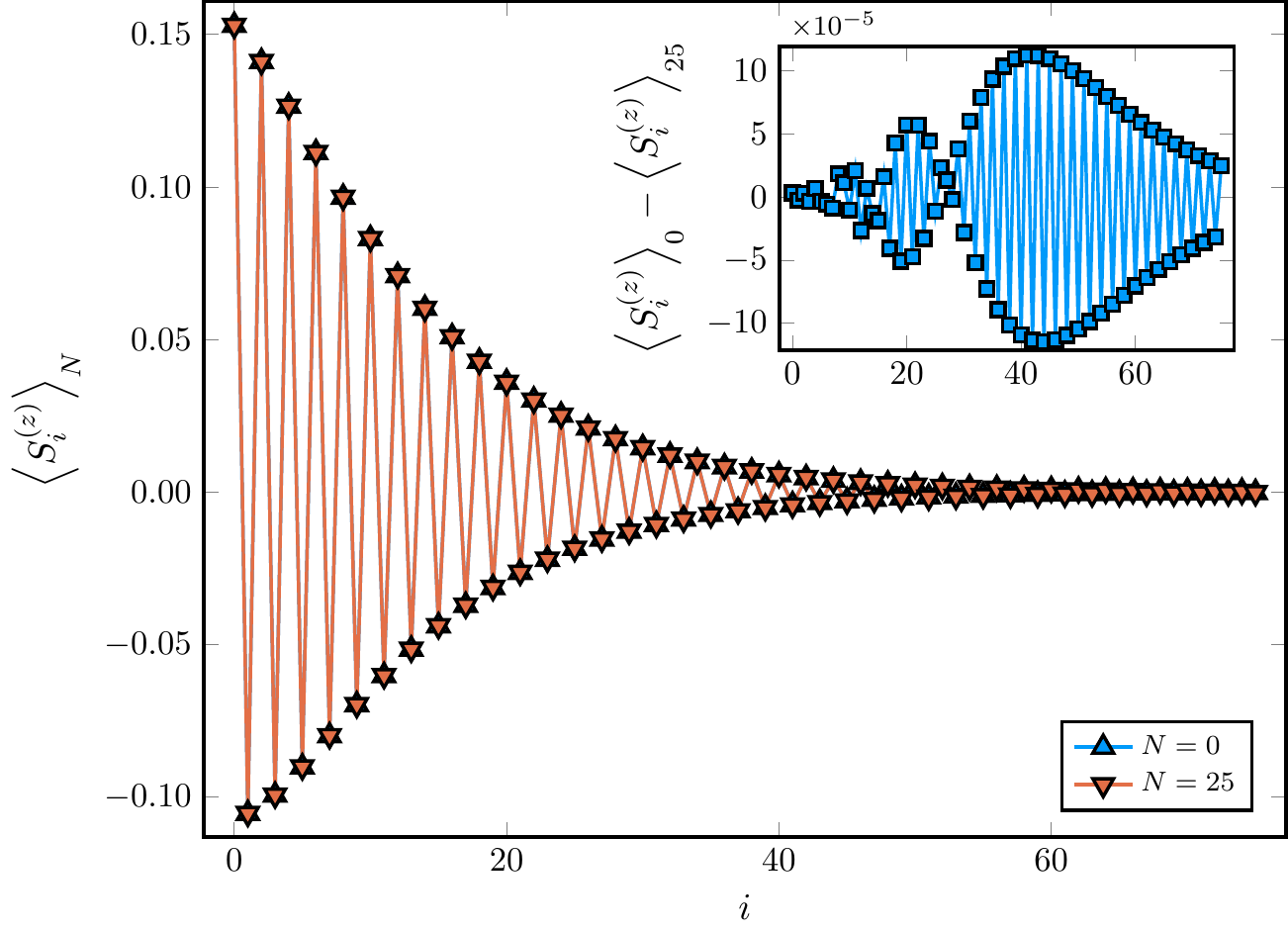}
\caption{%
The magnetization profile around the defect obtained by the MPS with $N=0$ and $N=25$. These two lines almost overlap. The inset shows the difference between the two profiles.%
}\label{fig:effspin}
\end{figure}

We plot the magnetization profiles of the state calculated for $N=0$ and $N=25$ in Fig.~\ref{fig:effspin}. 
The sum of the profile is equal to $1/2$, and therefore it can be regarded as an $S=1/2$ effective spin structure.
These two profiles agree well, and thus we can say that the effective spin structure is well represented by the MPS \eqref{eq:sMPS} even in the case of $N=0$.

\subsection{Trade-off between window length and precision}\label{sec:4N}

In the previous section, we treated the window length $N$ as a control parameter of the numerical calculation. Although the $N=0$ wavefunction gives a good approximate state, $N$ dependence is still an important matter. In this subsection, we study how the difference between $\ket{\Psi(A;\{B_{[i]}\}_{i=-\infty}^\infty)}$ and $\ket{\Psi(A;\{B_{[i]}\}_{i=-N}^N)}$ behaves as a function of $N$, where $\{B_{[i]}\}_{i=-N}^N$ denotes the set of $2N+1$ tensors optimized to minimize the energy for each window length.

\begin{figure}
\centering
\includegraphics[width=\linewidth]{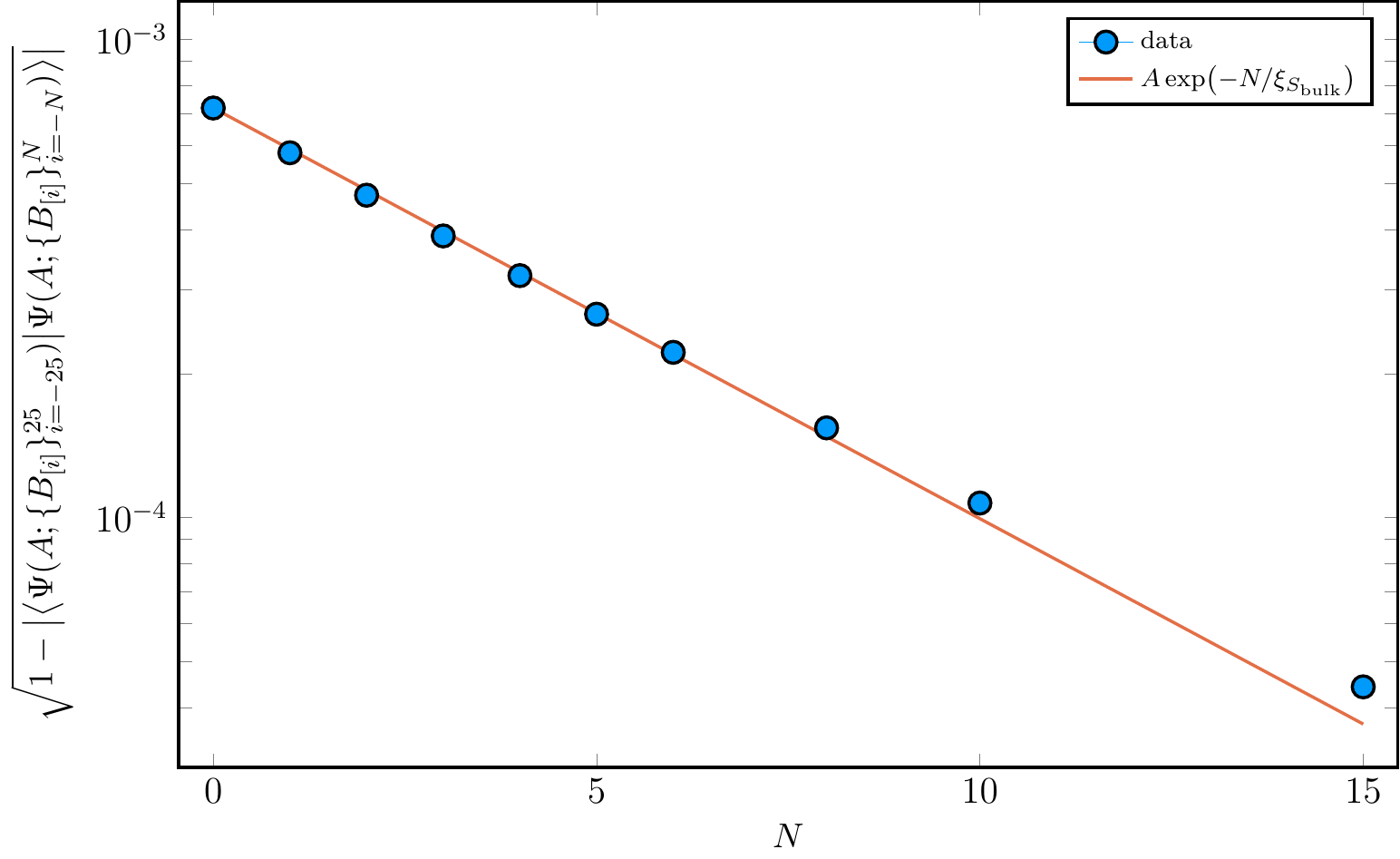}
\caption{%
The relation between the windows length $N$ and the fidelity defined by \eqref{eq:defdiff}. 
Blue circles denote the data of numerical calculation with $D=130$ and $N_{\mathrm{max}}=25$. 
The orange line denotes $A\exp(-N/\xi_{\mathrm{bulk}})$ line. 
The value of $A$ is chosen to fit the data at $N=0$.
}\label{fig:tradeoff}
\end{figure}

To study the $N$ dependence, we plot the fidelity
\begin{align}
    \sqrt{1-\abs*{\Braket{\Psi(\bar{A};\{\bar{B}_{[i]}\}_{i=-N_{\max}}^{N_{\max}})}{\Psi(A;\{B_{[i]}\}_{i=-N}^N)}}}\label{eq:defdiff}
\end{align}
in Fig.~\ref{fig:tradeoff}. 
We find that the fidelity \eqref{eq:defdiff} decreases with the correlation length of the bulk as $\sim\exp(-N/\xi_{\mathrm{bulk}})$. 

We believe that this behavior is general and does not depend on the detail of the model. 
We give an analytical result supporting this conjecture in Appendix \ref{sec:WL}.

\subsection{States with two localized spins}

Now, we study the case with two effective spins in the ABAHC.
The lattice structure is schematically drawn as
\begin{align*}
\includegraphics{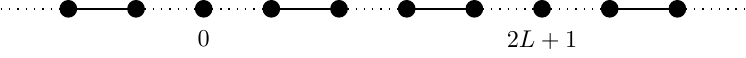}.
\end{align*}

Because the MPS \eqref{eq:sMPS} for $N=0$ already approximates the single effective spin state well, we construct a ``man-made'' state of two effective spins as
\begin{align}
    &\ket{\Psi^{(2)}_L(A;\{B\})}=\sum_{\{s\}} \vbf_\Lrm^\dagger \bigl(\prod_{m<0} A^{s_{2m},s_{2m+1}}\bigl)B^{s_0}\times \label{eq:handmade}\\ &\bigl(\ \prod_{\mathclap{0<m\leq L}}\  A^{s_{2m-1},s_{2m}}\bigl) B^{s_{2L+1}} \bigl(\prod_{L<m} A^{s_{2m},s_{2m+1}}\bigl) \vbf_\Rrm \ket{\{s\}},\nonumber
\end{align}
where $\{B\}$ denotes the central tensor in \eqref{eq:sMPS} for $N=0$. This state corresponds to a triplet state since two effective spins point in the same direction.

\begin{figure}
\centering
\resizebox{\linewidth}{!}{\includegraphics{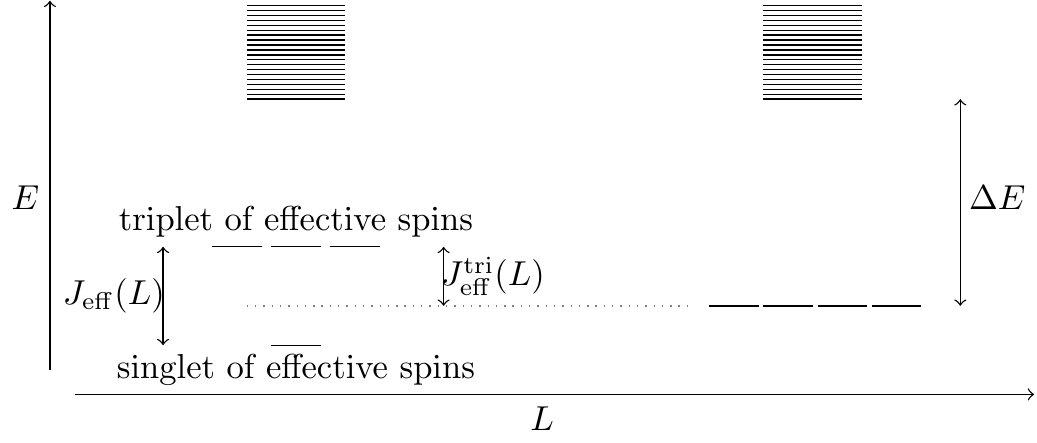}}
\caption{Schematic picture of the energy spectrum of two effective spins system. The effective exchange interaction breaks the degeneracy of the ground state.\label{fig:defeffint}}
\end{figure}

In the ABAHC, effective spins interact with each other, and this interaction breaks the degeneracy of the ground states, as illustrated in Fig.~\ref{fig:defeffint}.
We note that the exact degeneracy of the effective spin states \eqref{eq:multiaklt} originates from the frustration-free property of the AKLT Hamiltonian. 
By making use of \eqref{eq:handmade}, we study the effective interaction as a function of the distance between two defects.
We derive $J^{\text{tri}}_{\text{eff}}(L)$ as
\begin{align}
    &J_{\text{eff}}^{\text{tri}}(L) = E(L) - E(\infty) \sim \exp(-L/\xi_{\mathrm{bulk}})
\label{eq:effint}
\end{align}
by defining
\begin{align}
 E(L) = \frac{\Mel{\Psi^{(2)}_L(\bar{A};\{\bar{B}\})}{\Hcal}{\Psi^{(2)}_L(A;\{B\})}}{\Braket{\Psi^{(2)}_L(\bar{A};\{\bar{B}\})}{\Psi^{(2)}_L(A;\{B\})}}. \label{eq:defofE}
\end{align}
The detail derivation of \eqref{eq:effint} is given in Appendix \ref{sec:detailinteraction}.

\begin{figure}
\centering
\includegraphics[width=\linewidth]{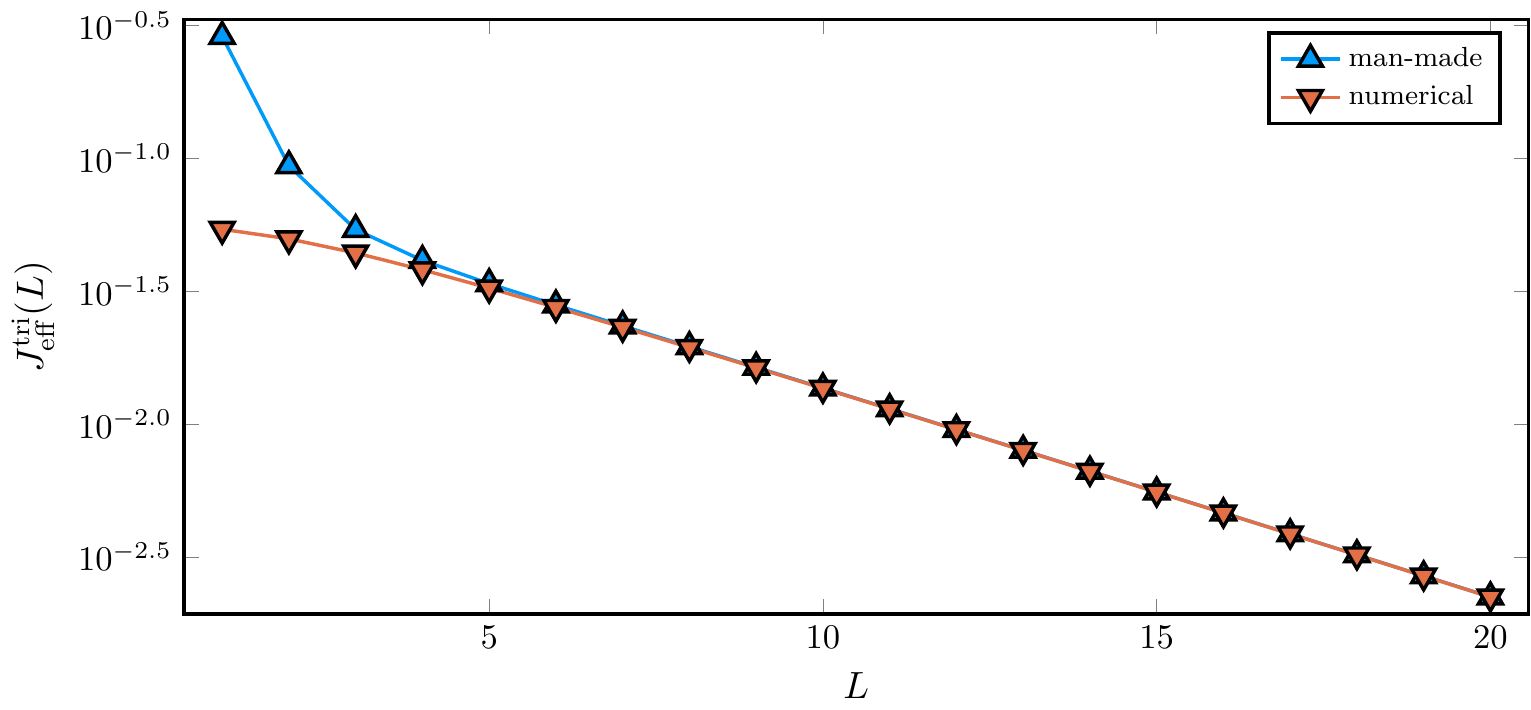}
\caption{The strength of the effective interaction $J^{\text{tri}}_{\text{eff}}$ as a function of the distance between two effective spins. Blue circles represent the energies calculated by the man-made state \eqref{eq:handmade}. 
After constructing man-made states, we optimize the states by the TDVP algorithm. In the optimization, we apply a uniform magnetic field $h_z$ satisfying $J_{\text{eff}} <h_z < \Delta E \approx 0.145$. Orange dots are the optimized energy. When $L$ larger than $5\approx\xi_{\mathrm{bulk}}$, the man-made state reproduces the effective interaction energy well.\label{fig:Jeff2}}
\end{figure}

Unlike the case of the AKLT model, the effective Hamiltonian acting on the site 0 (the detail definition is given in  \eqref{eq:heffbahc}) is modified by the existence of $B^{s_{2L+1}}$, and the same happens on the site $2L+1$. 
Therefore, even when the state $\ket{\Psi(A;\{B\})}$ can represent the ground state with high accuracy, the accuracy of the man-made state may become worse when the distance between two effective spins is not large enough.
We check the validity of \eqref{eq:handmade} by a numerical calculation. 
In Fig.~\ref{fig:Jeff2}, we plot $J^{\text{tri}}_{\text{eff}}(L)$ calculated by the man-made state \eqref{eq:handmade} and by the state optimizing the whole tensors in the window $[0,2L+1]$.
In the case $L > 5\approx \xi_{\mathrm{bulk}}$, the interaction was well reproduced by the man-made state.

Thus, we conclude that the MPS based characterization of the effective spins is useful for very general cases.

\section{Summary and Discussion}\label{sec:Summary}

We have studied localized effective spins induced by inhomogeneous lattice structures in gapped quantum spin systems.
As a prototype of such structure, first, we studied the AKLT model with doped $S=3/2$ spins.
We constructed the exact MPS representation of the ground state and analyzed the response to external magnetic fields.
We found that the response is given by a form of summation of local fields. 
Thus, by tuning the distribution of fields, we approximately manipulate the spins independently. 
However, if we take into account the non-orthonormality of the states, the operation interferes with each other, and the control is no more independent. 
We found that, for the case of two effective spins, we can construct qubit states which can be manipulated independently.
But, we also found such construction is impossible for the cases of more than two spins.

As a realistic model, we studied the ABAHC with defects, which has been studied experimentally, e.g., the work of Bertaina, \textit{et al.}~\cite{Bertaina2014-ww}.
The uniform MPS with impurity tensors can well approximate the ground state of this model as well as the case of the AKLT model.
But, some qualitative differences, due to the absence of the frustration-free property, exists.
The precision of the MPS approximation depends on the window length of the impurity tensors. 
We discover that this dependence is dominated by the bulk correlation length $\xi_{\mathrm{bulk}}$. 
We also studied the strength of the effective exchange interaction as a function of the separation of impurities, which was found to become small exponentially with the correlation length $\xi_{\mathrm{bulk}}$. 
For studying these characteristics, the MPS based characterization works well. 

In the ESR experiment~\cite{Bertaina2014-ww}, they found a sharp resonant peak, which is considered to be attributed to the effective spins.
Besides the sharp peak, they also found a broad structure, which should be attributed to fast motion, including the excited state.
In the present paper, we confined ourselves in the states below the gap.
In order to explain the experimental results, we have to take the excited states into account.
To study the effects of the excited states on the dynamics of the effective spins is a future work.

\section*{Acknowlegement}
The authors thank Prof. Tomotoshi Nishino and Prof. Hosho Katsura for fruitful discussion. 
The present work was supported by Grants-in-Aid for Scientific Research C (No.18K03444) from MEXT of Japan, and the Elements Strategy Initiative Center for Magnetic Materials (ESICMM:Grant Number 12016013) under the outsourcing project of MEXT. 
The authors also thank the Supercomputer Center, the Institute for Solid State Physics, the University of Tokyo, for the use of the facilities. 
H.N. was supported by Advanced Leading Graduate Course for Photon Science (ALPS), the University of Tokyo.

\onecolumngrid
\appendix

\section{TDVP algorithm and Analysis of the length scale of the window}\label{sec:WL}

In this appendix, we introduce the optimization algorithm for the tensors in the window, e.g., $\{B_{[i]}\}_{i=-N}^N$ in \eqref{eq:sMPS}. This algorithm is based on the imaginary time evolution and called time-dependent variational principle (TDVP).

Now, we consider $S=(d-1)/2$ spin chain with the Hamiltonian
\begin{align}
    \Hcal = \sum_{i\in\mathbb{Z}} \hat{h}_{i,i+1} +( \hat{h}^\loc_{-1,0} + \hat{h}^\loc_{0,1})  = \Hcal^{\mathrm{uniform}} + \Hcal^\loc_{[-1,1]}.\label{eq:huniloc}
\end{align}

As a starting point, we consider the state with the smallest window size:
\begin{align}
    \ket{\Psi(A;\{B\})} = \sums \vbf_\Lrm^{\dagger} \bigl(\prod_{n < 0} A^{s_{n}}\bigl)B^{s_0} \bigl(\prod_{0 < n} A^{s_{n}}\bigl) \vbf_\Rrm \ket{\{s\}},
\end{align}
where $A$ denotes the tensor which was calculated for the uniform Hamiltonian $\Hcal^{\mathrm{uniform}}$. The spectral decomposition of the transfer matrix $\Tcal_A^A = \sum_s A^s \otimes \bar{A}^s$ is given by
\begin{align}
    \Tcal_A^A = |r_1)(l_1| + \sum_{i=2}^{D^2} \lambda_i |r_i)(l_i|,
\end{align}
where $(l_i|$ and $|r_j)$ denote the left and right eigenvectors of $\Tcal_A^A$, respectively, satisfying $(l_i|r_j) = \delta_{i,j}$. Here, $1 > \abs{\lambda_2}\geq \abs{\lambda_3}\cdots$ is assumed. We also define the assosiated matrices $l_i$ and $r_j$ which fulfill
\begin{align}
    (l_i|\Tcal_A^A|r_j) = \sum_{s,b,b',k,k'} (l_i)_{b,k} A^s_{k,k'} (r_j)_{k',b'} \bar{A}^s_{b,b'}.
\end{align}

$B$ denotes the optimized tensor to minimize the total energy.  This optimization can be done by defining effective Hamiltonian and solve the eigenvalue problem. We define an operator transfer matrix as
\begin{align}
    \Jcal_{h}^{XY} = \sum_{s,t} \mel{s't'}{\hat{h}}{st} (X^s Y^t)\otimes (\bar{X}^{s'}\bar{Y}^{t'})
\end{align}
and shift the origin of the energy as $\hat{h} = \hat{h} - (l_1|\Jcal_h^{AA}|r_1)$.
Then, the norm and the matrix elements of $\Hcal$ are given by
\begin{align}
    &\braket{\Psi(\bar{A};\{\bar{B}\})}{\Psi(A;\{B\})} = (l_1| \Tcal_B^B |r_1) =: \bar{B}^{s'}_{l',r'} (N_\mathrm{eff})_{(s',l',r'),(s,l,r)} B^s_{l,r}
\end{align}
\begin{align}
    &\mel{\Psi(\bar{A};\{\bar{B}\})}{\Hcal}{\Psi(A;\{B\})} \nonumber\\ &=(l_1|\Jcal_{h}^{AA}\sum_{n=0}^\infty (\Tcal_A^A)^n \Tcal_B^B |r_1) + (l_1|\Jcal_{h+h^\loc}^{AB}|r_1) + (l_1|\Jcal_{h+h^\loc}^{BA}|r_1) + (l_1|\Tcal_B^B \sum_{n=0}^\infty (\Tcal_A^A)^n \Jcal_h^{AA} |r_1)\\
    &= \underbrace{(l_1|\Jcal_h^{AA}\sum_{k=2}^{D^2}\frac{1}{1-\lambda_k}|r_k)(l_k|}_{=:(\mathrm{LIBC}|}\Tcal_B^B|r_1)+(l_1|\Jcal_{h+h^\loc}^{AB}|r_1) + (l_1|\Jcal_{h+h^\loc}^{BA}|r_1) + (l_1|\Tcal_B^B \underbrace{ \sum_{k=2}^{D^2} \frac{1}{1-\lambda_k} |r_k)(l_k|\Jcal_{h}^{AA}|r_1)}_{=:|\mathrm{RIBC})}\\
    &=: \bar{B}^{s'}_{l',r'} (H_{\mathrm{eff}})_{(s',l',r'),(s,l,r)}B^s_{l,r}.\label{eq:heffbahc}
\end{align}
$(\mathrm{LIBC}|$ and $|\mathrm{RIBC})$ are called infinite boundary conditions~\cite{Phien2012-sq,Lo2019-dg,Michel2010-qm}. The optimized tensor $B$ is obtained by solving the generalized eigenvalue problem $\lambda N_{\mathrm{eff}} \vec{x} = H_{\mathrm{eff}} \vec{x}$.

Hereafter, for simplicity, we regard $\ket{\Psi(A,\{B\})}$ as a ``vacuum'' state and represent it by the following shorthand notation
\begin{align}
    &\ket{\ulin}  =  \ket{\Psi(A;\{B\})}\label{eq:N0sMPS}.
\end{align}
When some tensors in $\ket{\ulin}$ are replaced, we only denote the replaced tensors as
\begin{align}
    \ket{\ulin Y^{s_i}\ulin Z^{s_j}\ulin} = \sums \vbf_\Lrm^\dagger \bigl(\prod_{n < 0} A^{s_{n}}\bigl)B^{s_0} \bigl(\prod_{0 <n < i} A^{s_{n}}\bigl) Y^{s_i}\bigl(\prod_{i <n < j} A^{s_{n}}\bigl)Z^{s_j}\bigl(\prod_{j <n } A^{s_{n}}\bigl)\vbf_\Rrm \ket{\{s\}}.
\end{align}

Now, we consider the infinitesimal imaginary time evolution starting from $\ket{\ulin}$.
For small $\Delta\tau$, this evolution is obtained by approximating
\begin{align}
    e^{-\Delta \tau\Hcal}\ket{\ulin}=\ket{\ulin} - \Delta \tau \Hcal\ket{\ulin}
\end{align}
by
\begin{align}
    \ket{\ulin (A^{s_{-N}}+\Delta\tau C_{[-N]}^{s_{-N}}) \cdots (A^{s_{N}}+\Delta\tau C_{[N]}^{s_{N}}) \ulin} = \ket{\ulin} + \Delta\tau \sum_{-N\leq i \leq N}\ket{\ulin C_{[i]}^{s_i}\ulin}.
\end{align}
In order to calculate this time evolution, we solve the minimization problem
\begin{align}
    \{C_{[i]}\}_{i=-N}^N=&\argmin_{\{\tilde{C}_{[i]}\}_{i=-N}^N} \norm*{\Hcal\ket{\ulin} + \sum_i\ket{\ulin \tilde{C}_{[i]}^{s_i}\ulin}}^2\\=& \argmin_{\{\tilde{C}_{[i]}\}_{i=-N}^N} \left(\sum_{-N\leq i \leq N} \left(\norm{\ket{\ulin \tilde{C}_{[i]}^{s_i}\ulin}}^2+\mel{\ulin \bar{\tilde{C}}_{[i]}^{s_i}\ulin}{\Hcal}{\ulin} + \mel{\ulin}{\Hcal}{\ulin \tilde{C}_{[i]}^{s_i}\ulin}\right)\right)\nonumber.
\end{align}
However, because of the gauge degrees of freedom, i.e., $\ket{\ulin C_{[i]}^{s_i}\ulin} + \ket{\ulin C_{[i+1]}^{s_{i+1}}\ulin} = \ket{\ulin (C_{[i]}^{s_i} + A^{s_i}X)\ulin} + \ket{\ulin (C_{[i+1]}^{s_{i+1}} - XA^{s_{i+1}})\ulin}$ for arbitrary $D\times D$ matrix $X$, $\{C_{[i]}\}$ is not uniquely determined. Then, we consider the following constrained optimization problem:
\begin{align}
    \{C_{[i]}\}_{i=-N}^N=& \argmin_{\{\tilde{C}_{[i]}\}_{i=-N}^N} \left(\sum_{-N\leq i \leq N} \left(\norm{\ket{\ulin \tilde{C}_{[i]}^{s_i}\ulin}}^2+\mel{\ulin \bar{\tilde{C}}_{[i]}^{s_i}\ulin}{\Hcal}{\ulin}+\mel{\ulin}{\Hcal}{\ulin \tilde{C}_{[i]}^{s_i}\ulin}\right)\right)\label{eq:tdvpdef1}\\ &\text{subject to}\ \Braket{\ulin}{\ulin \tilde{C}_{[i]}^{s_i} \ulin} =0\ \text{for}\ i\neq 0\ \text{and}\ \Braket{\ulin \bar{\tilde{C}}_{[j]}^{s_j} \ulin}{\ulin \tilde{C}_{[i]}^{s_i} \ulin} = 0\ \text{for}\ i\neq j.\nonumber
\end{align}

By introducing a $d\times D\times D(d-1)$ tensor $(V_L)^s_{a,a'}$ satisfying
\begin{align}
    (\sqrt{l_1}|\Tcal_A^{V_L} = 0,\quad (\mathbb{I}_{D\times D}|\Tcal_{V_L}^{V_L} = (\mathbb{I}_{D(d-1)\times D(d-1)}|\label{eq:vldef}
\end{align}
and defining a parameter representation of $C^{s_i}$ as
\begin{align}
    C(X)^{s_i}_{a,a'} = \sum_{b,b',b''} (\sqrt{l_1}^{-1})_{a,b} (V_L)^{s_i}_{b,b'} X_{b',b''} (\sqrt{r_1}^{-1})_{b'',a'},
\end{align}
the constraints in \eqref{eq:tdvpdef1} automatically satisfied for $i<0$ sites because $(l_1|\Tcal_{A}^{C(X)} = (\sqrt{l_1}|\Tcal_A^{V_L X \sqrt{r_1}^{-1}} = 0$. Here, $X$ denotes a $D(d-1)\times D$ matrix. Furthermore, the norm of $\ket{\ulin C(X)^{s_i}\ulin}$ is given by a simple form as
\begin{align}
    \braket{\ulin \overline{C(X)}^{s_i}\ulin}{\ulin C(X)^{s_i}\ulin} = (l_1|\Tcal_{C(X)}^{C(X)}|r_1) = (\mathbb{I}_{D\times D}|\Tcal_{V_L X}^{V_L X}|\mathbb{I}_{D\times D}) = (\mathbb{I}_{D(d-1)\times D(d-1)}|\Tcal_X^X|\mathbb{I}_{D\times D}) = \Tr(XX^\dagger).
\end{align}
We construct $V_R$ and define $C(X)^{s_i}$ for $i > 0$ in the same manner. Then, we can solve \eqref{eq:tdvpdef1} for every sites independently:
\begin{align}
    C_{[0]}^{s_0} &= \argmin_{\tilde{C}}\left( (l_1|\Tcal_{\tilde{C}}^{\tilde{C}}|r_1) +\mel{\ulin \bar{\tilde{C}}^{s_0}\ulin}{\Hcal}{\ulin} + \mel{\ulin}{\Hcal}{\ulin \tilde{C}^{s_0}\ulin}\right)\\
    C_{[i\neq0]}^{s_i} &= C(X_{[i]})^{s_i}\\
    X_{[i\neq0]} &= \argmin_{\tilde{X}}\left( \Tr(\tilde{X}\tilde{X}^\dagger)+\mel{\ulin C(\bar{\tilde{X}})^{s_i}\ulin}{\Hcal}{\ulin} + \mel{\ulin}{\Hcal}{\ulin C(\tilde{X})^{s_i}\ulin} \right).\label{eq:optimizex}
\end{align}

By repeating this step, we simulate the imaginary time evolution starting from $\ket{\ulin}$ as 
\begin{align}
 \ket{\ulin B^{s_{-N}}(\tau)\cdots B^{s_{N}}(\tau)\ulin} = e^{-\tau \Hcal}\ket{\ulin}/\norm{e^{-\tau \Hcal}\ket{\ulin}}
\end{align}
for every $N$.

The result shown in Fig.~\ref{fig:tradeoff} suggests that
\begin{align}
    1 - \abs*{\Braket{\bar{B}^{s_{-\infty}}(\infty)\cdots \bar{B}^{s_{\infty}}(\infty)}{\ulin B^{s_{-N}}(\infty)\cdots B^{s_{N}}(\infty)\ulin}} \sim  \abs{\lambda_2}^{2N}.\label{eq:toprove}
\end{align}
We analyze the short time behavior, where the dynamics can be regarded as linear, of the left hand side of \eqref{eq:toprove}, and prove that
\begin{align}
    1 - \abs*{\Braket{\bar{B}^{s_{-\infty}}(\Delta \tau)\cdots \bar{B}^{s_{\infty}}(\Delta \tau)}{\ulin B^{s_{-N}}(\Delta \tau)\cdots B^{s_{N}}(\Delta \tau)\ulin}} \sim  \Delta\tau^2\abs{\lambda_2}^{2N}\label{eq:infsml}
\end{align}
is satisfied if $\Delta\tau$ is enought small.
Because Fig.~\ref{fig:tradeoff} suggests the initial state $\ket{\ulin}$ is close to the $\tau=\infty$ state, we assume that the linear dynamics \eqref{eq:infsml} gives the dominant contribution of \eqref{eq:toprove}.

We define $\epsilon_m = \begin{cases}
    (l_1|\Tcal_{C_{[0]}}^{C_{[0]}}|r_1) & (m=0) \\ \Tr(X_{[m]}X_{[m]}^\dagger) &(m\neq 0) 
    \end{cases}$, and then the left hand side of \eqref{eq:infsml} is written as
\begin{align}
    &1-\frac{1 + \Delta\tau^2\sum_{-N\leq m \leq N}\epsilon_m}{\sqrt{1 + \Delta\tau^2\sum_{-\infty < m < \infty}\epsilon_m}\sqrt{1 + \Delta\tau^2\sum_{-N \leq m \leq N}\epsilon_m}}\\
    =&1-\left(1 + \Delta\tau^2\sum_{-N\leq m \leq N}\epsilon_m\right)\left(1 -\frac{1}{2} \Delta\tau^2\sum_{-\infty < m < \infty}\epsilon_m\right)\left(1 -\frac{1}{2} \Delta\tau^2\sum_{-N \leq m \leq N}\epsilon_m\right)\\
    =&\frac{1}{2}\Delta\tau^2\sum_{\abs{m}>N}\epsilon_m.\label{eq:shorttime}
\end{align}

$X_{[m]}$ is obtained by taking ${\partial}/{\partial \bar{\tilde{X}}}$ of the right hand side of \eqref{eq:optimizex}. For simplicity, we assume $m < -2$ and define
\begin{align}
    &\ket{a}_\Lrm = \sum_{\{s_i\}_{i\leq m+1}} \left[\vbf_\Lrm^\dagger\bigl(\prod_{n \leq m+1}A^{s_n}\bigl) \right]_{1,a} \ket{\{s_i\}_{i\leq m+1}}\\
    &\ket{a}_\Rrm = \sum_{\{s_i\}_{m+1<i}} \left[\bigl(\prod_{m+1<n<0}A^{s_n}\bigl)B^{s_0}\bigl(\prod_{0<n} A^{s_n}\bigl)\vbf_\Rrm \right]_{a,1} \ket{\{s_i\}_{m<i}}\\
    &\ket{a}_{\Lrm'\Lrm} = \sum_{b,b'}\sum_{\{s_i\}_{i\leq m}} \left[\vbf_\Lrm^\dagger \bigl(\prod_{i<m} A^{s_i} \bigl)\right]_{1,b} (l_1^{-1/2})_{b,b'} (V_L)^{s_m}_{b',a} \ket{\{s_i\}_{i\leq m}}\\
    &\ket{a,a'}_{\Lrm'\mathrm{C}} = \sum_{b,s_{m+1}} (r_1^{-1/2})_{a,b}A^{s_{m+1}}_{b,a'}\ket{s_{m+1}}.
\end{align}
Since
\begin{align}
    &\ket{\Psi(A;\{B\})} = \sum_a \ket{a}_\Lrm\otimes\ket{a}_\Rrm\\
    &\ket{\ulin C(\tilde{X})^s \ulin} = \sum_{a,a',a''} \ket{a}_{\Lrm'\Lrm}\otimes \tilde{X}_{a,a'}\ket{a',a''}_{\Lrm'\mathrm{C}}\otimes\ket{a''}_\Rrm,
\end{align}
$X_{[m]}$ is obtained as
\begin{align}
    (X_{[m]})_{a,a'}=\left(\sum_{a''}\bra{a}_\Lrm\otimes\bra{a',a''}_\mathrm{C}\otimes\bra{a''}_\Rrm\right)\Hcal\ket{\Psi(A;\{B\})}.
\end{align}

Because of the property \eqref{eq:vldef}, if an operator $\mathcal{O}$ acts only on the right-hand side sites of site $m$, $\left(\sum_b\bra{a}_\Lrm\otimes\bra{a',b}_\mathrm{C}\otimes\bra{b}_\Rrm\right)\mathcal{O}\ket{\Psi(A;\{B\})}$ vanishes. We now define an $m$ independent matrix
\begin{align}
    (Y_{a,a'})_{b,b'} = \left(\bra{a}_{\Lrm'\Lrm}\otimes\bra{a',b}_{\Lrm'\mathrm{C}}\right)\left(\sum_{i\leq m+1} h_{i-1,i}\right)\ket{b'}_\Lrm,
\end{align}
and then $X_{[m]}$ can be written as
\begin{align}
    (X_{[m]})_{a,a'} &= (Y_{a,a'}|(\Tcal_A^A)^{\abs{m}-2}\Tcal_B^B |r_1)\\
    &= \underbrace{(Y_{a,a'}|r_1)}_{=0}\underbrace{(l_1|\Tcal_B^B|r_1)}_{=1} + \sum_{i=2}^{D^2} \lambda_i^{\abs{m}-2}(Y_{a,a'}|r_i)(l_i|\Tcal_B^B|r_1) \sim \lambda_2^{\abs{m}}\eta_{a,a'}.\label{eq:asy}
\end{align}

\eqref{eq:infsml} is given as a consequence of \eqref{eq:shorttime} and \eqref{eq:asy}.

\section{Derivation of the asymptotic form of the interaction}\label{sec:detailinteraction}

We consider the Hamiltonian with two inhomogeneities
\begin{align}
    \Hcal = \Hcal^{\mathrm{uniform}} + \Hcal^\loc_{[-1,1]} + \Hcal^\loc_{[L,L+2]},
\end{align}
and the man-made state
\begin{align}
    \ket{\Psi_L^{(2)}(A;\{B\})} = \ket{\cdots A^{s_{-1}}B^{s_0}A^
    {s_1}\cdots A^{s_{L}}B^{s_{L+1}}A^{s_{L+2}}\cdots}.
\end{align}
We define $\Jcal^\loc = \Jcal_{h+h^\loc}^{AB}\Tcal_A^A + \Tcal_A^A\Jcal_{h+h^\loc}^{BA}$ and shift the origin of $h^\loc$ to satisty
\begin{align}
    0 = \mel{\Psi(\bar{A};\{\bar{B}\})}{\Hcal^{\mathrm{uniform}} +\Hcal^\loc_{[-1,1]}}{\Psi(A;\{B\})} = (\mathrm{LIBC}|\Tcal_B^B|r_1) + (l_1|\Tcal_B^B|\mathrm{RIBC}) + (l_1|\Jcal^\loc|r_1).
\end{align}

The denominator and numerator of $E(L)$ are given as
\begin{align}
    \braket{\Psi_L^{(2)}(\bar{A};\{\bar{B}\})}{\Psi_L^{(2)}(A;\{B\})} &= (l_1|\mathcal{T}_B^B (\mathcal{T}_A^A)^L \mathcal{T}_B^B|r_1)  = (l_1|\mathcal{T}_B^B|r_1)(l_1|\mathcal{T}_B^B|r_1) + \sum_{i=2}^{D^2}\lambda_i^L(l_1|\mathcal{T}_B^B|r_i)(l_i| \mathcal{T}_B^B|r_1)\\
    &= 1 + \mathcal{O}(\exp(-L/\xi_{\mathrm{bulk}})).
\end{align}
and
\begin{align}
&\mel{\Psi_L^{(2)}(\bar{A};\{\bar{B}\})}{\mathcal{H}}{\Psi_L^{(2)}(A;\{B\})}\\
    =\ &(\text{LIBC}|\mathcal{T}_B^B (\mathcal{T}_A^A)^L \mathcal{T}_B^B |r_1)+ (l_1|\Jcal^\loc (\mathcal{T}_A^A)^{L-1} \mathcal{T}_B^B |r_1)+ (l_1| \mathcal{T}_B^B (\mathcal{T}_A^A)^{L-1} \Jcal^\loc|r_1) + (l_1| \mathcal{T}_B^B (\mathcal{T}_A^A)^L \mathcal{T}_B^B |\text{RIBC})\nonumber\\
    +& \sum_{i=0}^{L-2}(l_1|\mathcal{T}_B^B (\mathcal{T}_A^A)^i \mathcal{J}_{h}^{(AA)} (\mathcal{T}_A^A)^{L-i-2} \mathcal{T}_B^B |r_1)\\
    =\ &(\text{LIBC}|\mathcal{T}_B^B|r_1)\underbrace{(l_1|\mathcal{T}_B^B |r_1)}_{=1} + \sum_{i=2}^{D^2}\lambda_i^L(\text{LIBC}|\mathcal{T}_B^B|r_i)(l_i|\mathcal{T}_B^B |r_1) +(l_1|\Jcal^\loc |r_1)\underbrace{(l_1|\mathcal{T}_B^B |r_1)}_{=1}+ \sum_{i=2}^{D^2} \lambda_i^{L-1}(l_1|\Jcal^\loc |r_i)(l_i| \mathcal{T}_B^B |r_1)\nonumber\\
    +& \underbrace{(l_1|\mathcal{T}_B^B |r_1)}_{=1}(l_1| \Jcal^\loc|r_1) + \sum_{i=2}^{D^2}\lambda_i^{L-1}(l_1| \mathcal{T}_B^B|r_i)(l_i| \Jcal^\loc|r_1) + \underbrace{(l_1|\mathcal{T}_B^B |r_1)}_{=1}(l_1| \mathcal{T}_B^B |\text{RIBC}) + \sum_{i=2}^{D^2}\lambda_i^L(l_1| \mathcal{T}_B^B |r_i)(l_i| \mathcal{T}_B^B |\text{RIBC})\nonumber\\
    +& \sum_{j=2}^{D^2}\sum_{i=0}^{L-2}\left(\lambda_j^{L-i-2}\underbrace{(l_1|\mathcal{T}_B^B |r_1)}_{=1}(l_1| \mathcal{J}_{h}^{AA}|r_j)(l_j|  \mathcal{T}_B^B |r_1) + \lambda_j^i(l_1|\mathcal{T}_B^B |r_j)(l_j| \mathcal{J}_{h}^{AA}|r_1)\underbrace{(l_1|\mathcal{T}_B^B |r_1)}_{=1}\right)\nonumber\\
    +&\sum_{j=2}^{D^2}\sum_{k=2}^{D^2}\sum_{i=0}^{L-2} \lambda_j^{i} \lambda_k^{L-i-2}(l_1|\mathcal{T}_B^B |r_j)(l_j| \mathcal{J}_{h}^{AA}|r_k)(l_k|  \mathcal{T}_B^B |r_1)\\
    =\ & 2\left(\underbrace{(\text{LIBC}|\mathcal{T}_B^B|r_1)+(l_1|\mathcal{T}_B^B|\text{RIBC}) + (l_1|\Jcal^\loc|r_1)}_{=0}\right)
    + \sum_{i=2}^{D^2}\frac{-\lambda_i^{L-1}}{1-\lambda_i}\left((l_1|\mathcal{J}_h^{AA}|r_i)(l_i|\mathcal{T}_B^B|r_1) + (l_1|\mathcal{T}_B^B|r_i)(l_i|\mathcal{J}_h^{AA}|r_1)\right)\nonumber\\ +&  \sum_{i=2}^{D^2}\lambda_i^{L}\left((\text{LIBC}|\mathcal{T}_B^B|r_i)(l_i|\mathcal{T}_B^B |r_1)+(l_1|\mathcal{T}_B^B |r_i)(l_i|\mathcal{T}_B^B|\text{RIBC})\right)\nonumber \\+& \sum_{i=2}^{D^2} \lambda_i^{L-1}\left( (l_1|\Jcal^\loc|r_i)(l_i|\mathcal{T}_B^B|r_1)+ (l_1|\mathcal{T}_B^B|r_i)(l_i|\Jcal^\loc|r_1)\right)
    +\sum_{j=2}^{D^2}\sum_{k=2}^{D^2}\sum_{i=0}^{L-2} \lambda_j^{i} \lambda_k^{L-i-2}(l_1|\mathcal{T}_B^B |r_j)(l_j| \mathcal{J}_{h}^{AA}|r_k)(l_k|  \mathcal{T}_B^B |r_1)\\
    =\ & \mathcal{O}(L\exp(-L/\xi_{\mathrm{bulk}})).
\end{align}

\section{Detail calculation of the AKLT model}\label{sec:detailAKLT}

By using the notation introduced in Appendix~\ref{sec:WL}, the transfer matrices of \eqref{eq:defA} and \eqref{eq:defB} are given as
\begin{align}
    &\Tcal_A^A = |r_1)(l_1| -\frac{1}{3}\left(|r_2)(l_2|+|r_3)(l_3|+|r_4)(l_4|\right)\\
    &\Tcal_{B_\uparrow}^{B_\uparrow} = \Tcal_A^A + \frac{1}{3}\left(|r_2)(l_1|-|r_1)(l_2|\right)\\
    &\Tcal_{B_\downarrow}^{B_\downarrow} = \Tcal_A^A - \frac{1}{3}\left(|r_2)(l_1|-|r_1)(l_2|\right)\\
    &\Jcal_{S_z}^A = \frac{2}{3}\left(|r_2)(l_1|-|r_1)(l_2|\right).
\end{align}

Now, we consider the case of $k=3$ of \eqref{eq:multiaklt}. We fix the positions of the doped spins as $j_1=0,\ j_2=L+1,\ j_3=L+L'+2$. To show the difference between $k=2$ and $k>2$, we consider
\begin{align}
    \mel{\uparrow_1,\uparrow_2,\uparrow_3}{S^z_{L+L'+L''+3}}{\uparrow_1,\uparrow_2,\uparrow_3}=(l_1|\Tcal_{B_\uparrow}^{B_\uparrow}(\Tcal_A^A)^L\Tcal_{B_\uparrow}^{B_\uparrow}(\Tcal_A^A)^{L'}\Tcal_{B_\uparrow}^{B_\uparrow}(\Tcal_A^A)^{L''}\Jcal_{S_z}^A|r_1).\label{eq:3scat}
\end{align}
Different from the case of $k=2$, \eqref{eq:3scat} contains theree times scattering term as
\begin{align}
    (l_1|\xrightarrow[\times-\frac{1}{3}]{\Tcal_B^B} (l_2|\xrightarrow[\times\left(-\frac{1}{3}\right)^L]{(\Tcal_A^A)^L}(l_2|\xrightarrow[\times\frac{1}{3}]{\Tcal_B^B} (l_1| \xrightarrow[\times 1]{(\Tcal_A^A)^{L'}}(l_1| \xrightarrow[\times -\frac{1}{3}]{\Tcal_B^B} (l_2| \xrightarrow[\times \left(-\frac{1}{3}\right)^{L''}]{(\Tcal_A^A)^{L''}} \underbrace{(l_2| \Jcal^A_{S_z} |r_1)}_{\frac{2}{3}} = -\frac{2}{3}\left(-\frac{1}{3}\right)^{L+L''+2}.\label{eq:mltsct}
\end{align}
Such terms resulting from many times scattering make it impossible to represent $\ubf_3^\dagger \Hcal'\ubf_3$ in a simple form as \eqref{eq:22hmat}.

\twocolumngrid
\bibliography{bibfile}

\end{document}